# A signature for the absence of event horizons

J. Barbieri
NAWC-WD, China Lake, CA 93555
G. Chapline
Lawrence Livermore National Laboratory, Livermore, CA 94550

**Abstract**
A sharp dip in the spectrum of γ rays coming from compact objects below 70 MeV would be an unambiguous signal that compact astrophysical objects have a physical surface, and there is no event horizon. Observation of this effect would open a window for the empirical study of Planck scale physics

## 1. Introduction

Black hole solutions of the classical Einstein equations pose a number of conceptual difficulties, not the least of which is incompatibility with elementary quantum mechanics. It has been suggested that in reality the interior of a compact object is a "squeezed" version of the ordinary space-time vacuum [1]. This led to the suggestion [2,3] that the surface separating the squeezed vacuum from the ordinary vacuum is a physical surface that produces observable effects rather than an event horizon. One of us (GC) has coined the name "dark energy star" for a compact object where this surface is a quantum critical layer [4]. In sharp contrast with the celebrated prediction of classical general relativity that nothing happens to particles as they fall through the event horizon, one finds that in the quantum criticality picture ordinary matter will undergo a dramatic transformation at the surface of a dark energy star. In particular, elementary particles whose initial momentum exceeds a characteristic value $Q_0$, on the order of $100(M_o/M)^{1/2}$ MeV/c, where M is the mass of the compact object, will be strongly scattered and can decay into other elementary particles [5]. These interactions are similar to those experienced by quasi-particles in a quantum liquid near to a quantum critical point. A remarkable prediction of this quantum criticality picture is that protons hitting the surface of a compact astrophysical object will decay into positrons and mesons [5].

It happens that the quarks and gluons inside nucleons typically have momenta that exceed $Q_0$ for all known compact objects, and therefore will undergo strong interactions at the surface if the surface is a quantum critical layer. In grand unified models of elementary particles such as the Georgi-Glashow SU(5) model [6] nucleon disappearance will then proceed via the baryon number violating reactions:

$$u(2/3) \rightarrow e(+1) + \bar{u}(-2/3) + \bar{d}(1/3)$$

$$d(-1/3) \rightarrow e(+1) + 2\bar{u}(-2/3)$$

(1)

where *u* and *d* are the "up" and "down" quarks found inside protons and neutrons. In the Georgi-Glashow model nucleon decay will primarily yield positrons and mesons. Under ordinary circumstances nucleon decay is highly suppressed by the very large masses of the intermediate bosons associated with baryon number violation, and in fact has never been observed in the laboratory. However, quarks falling into the quantum critical layer at the surface of a dark energy star acquire a rest mass approaching or exceeding the "grand unification" mass scale where baryon number violating processes are expected to be just as important as the elementary particle processes studied in earthbound accelerator experiments. As a consequence ordinary matter hitting the surface of the compact object will produce an outgoing flux of MeV positrons and γ rays.

In general the interactions near to a quantum critical point can be simply approximated as a universal four point interaction [7]. As in Fermi's theory of beta decay this will lead to 3-body decays with a universal spectrum. For a 3-body decay product emitted at an angle θ with respect to the normal to the surface of the compact object this universal spectrum has the form:

$$\frac{dN}{d\Omega d\omega} = \frac{27}{16\pi^2 \omega_0} \left[ 3\frac{\omega}{\omega_0} \left( 1 - 3\frac{\omega}{\omega_0} - 2\sqrt{\frac{\omega}{\omega_0}} \cos\theta \right) \right]^{1/2}, \qquad (2)$$

where $\hbar\omega_0$ is the initial energy of the elementary particle that is incident on the compact object and $\hbar\omega$ is the energy the decay product would have if it escaped to infinity. For the compact objects typically encountered in astronomy the quantum critical layer is quite thin compared to the Schwarzschild radius, so a distant observer will only be able to see decay products emitted in a direction nearly perpendicular to the surface (i.e. θ=0). The actual spectrum of a weakly interacting decay product, e.g. of a decay lepton, that would be seen by distant observer would of course depend on the distribution of initial energies $\hbar\omega_0$ for the incident elementary particle. In the case of nucleons incident on the surface this distribution would be the distribution of quark or gluon energies inside the incident nucleon. In a previous paper [8] the spectrum of positrons that would be produced by low energy nucleons falling onto the surface of the compact object was calculated using Eq. (2) and a distribution for the quark momenta inside a nucleon obtained by using the Altarelli-Parisi equations to extrapolate the quark distribution measured in laboratory accelerator experiments to the Planck energy.

Remarkably an excess of 511 keV radiation from positron annihilation has been observed coming from the vicinity of the center of our galaxy **[9,10,11],** which apparently is due to positron annihilation in the interstellar medium. Unfortunately, it is not possible to directly compare our predicted spectrum of positrons due to nucleon decay with the observations of positron annihilation radiation since positrons lose their energy in the interstellar medium. However, the absence of the in-flight positron annihilation γ-rays expected if the positrons had very high energies [12] is consistent with the spectrum of positrons predicted by Eq. (2). For example, if the positrons were the result of the decay of heavy dark matter particles they would be expected to energies > GeV rather than the MeV energies predicted by Eq. (2). On the other hand, because the positrons can diffuse some distance away from a compact object before annihilating, it difficult to say for sure that the observed positrons come from nucleon decay at the location of the compact object. Fortunately a more favorable opportunity for checking Eq. (2) may be provided by the $\pi^0$ decay γ-rays coming directly from the surface of the compact object.

γ-rays are only weakly absorbed by the interstellar medium, and therefore $\pi^0$ decay γ-rays produced at the surface of a compact object can tell us something about the nature of the surface. In particular the spectrum of $\pi^0$ decay γ-rays coming from the surface will tell us whether there is a quantum critical layer at the surface rather than an event horizon. The important point is that the kinematics of the γ-rays from nucleon decay [delete that we are allowed to see] is constrained by the presence of the quantum critical layer as illustrated in Fig 1. This kinematics is different from positron production in that the positrons that we see must have been emitted directly in the backward direction (θ=0), whereas consideration of the angle-dependence of the fluorescence spectrum (2) and solid angle factors implies that the $\pi^0$ mesons from nucleon decay are mostly emitted parallel to the surface of the compact object (θ near to π/2). Emission of γ-rays from $\pi^0$ containing jets directed inward (θ > π/2) is highly suppressed because the initiating anti-quark would be rapidly thermalized in the quantum critical layer. Therefore a dramatic feature of the $\pi^0$ decay γ-ray spectrum that is directly tied to the existence of the quantum critical layer will be a sharp cutoff in the spectrum below $m_\pi c^2/2$.

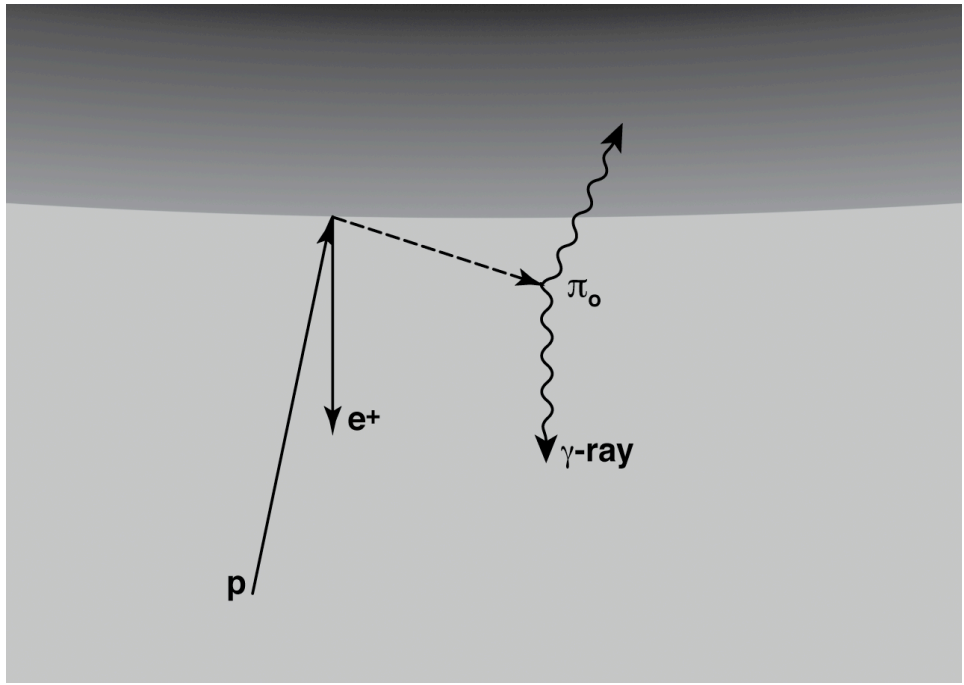

**Fig 1.** Kinematics of positron and $\pi^0$ decay γ-ray production near to the surface of a dark energy star

## 2. Calculation

As indicated in Eq. (1) an immediate result of an encounter of a nucleon with the quantum critical surface at the surface of a compact object is the production of anti-quarks. These anti-quarks will quickly become "dressed" as mesons due to quark confinement. Since this dressing occurs at an extremely high Tolman temperature, it is plausible that that a jet-like model can be used to describe the meson production. . For the purposes of calculating the spectrum of $\pi^0$ decay $\gamma$ rays we will adopt a "mini-jet" model where we assume that the hadronic jet produced by the decay anti-quark consists of a single $\pi^0$ whose momentum is equal to the momentum $\hbar\omega'/c$ of the decay anti-quark. That is we will set $\hbar\omega' = m_\pi c^2 \beta / \sqrt{(1-\beta^2)}$, where $c\beta$ is the $\pi^0$ velocity. The production cross-section (2) and solid angle factors favor the production of decay anti-quarks near to $\theta=\pi/2$, and hence the mini-jet $\pi^0$ s are produced predominately parallel to the surface of the compact object. The $\gamma$-rays we see must of course be emitted perpendicular to the surface (cf. Fig 1). Taking into account the transverse Doppler shift of the $\pi^0$ decay $\gamma$-ray this allows us to identify the $\gamma$ ray energy observed by a distant observer as

$$E_\gamma = \frac{\hbar\omega'}{2\beta}. \tag{3}$$

For small momenta this energy approaches $m_\pi c^2/2$. The spectrum for $\pi^0$ decay $\gamma$ rays escaping to infinity for $E_\gamma > m_\pi c^2/2$ will be given by

$$\frac{dN_\gamma}{dE_\gamma} = \frac{2}{\beta} \int_{\omega_{min}}^{\infty} \frac{d\omega}{\omega} F(\omega',\omega) \int_{\omega}^{\infty} \frac{dN_p}{dE_p} \frac{q(x,Q^2)}{E_p} dE_p, \tag{4}$$

where $\omega_{min}$ is either $3\omega'$ or the cutoff frequency $\omega_c \equiv Q_0 c/\hbar$ depending on which is larger, $x=\hbar\omega/E_p$, $dN_p/dE_p$ is the proton spectrum, $q(x,Q^2)$ is a normalized distribution for the quark momenta inside a nucleon, and $F(\omega',\omega)$ is the fluorescence spectrum (2) for $\theta=\pi/2$. According to Liu, et.al. [13] the energy spectrum of protons near to a compact object will have a the form:

$$\frac{dN_p}{dE}(E) = F_o(E_0/E)^\alpha. \tag{5}$$

By changing the integration variables in eq. (4) to x and $\omega'/\omega$ one can show that for a proton spectrum of the form (5) the spectrum (4) can be written in the form:

$$\frac{dN_\gamma}{dE_\gamma} = \frac{2F_0}{\beta} <(\frac{\omega'}{\omega})^{\alpha-1}><(\frac{\omega}{E_p})^{\alpha-1}>(\frac{E_0}{E\gamma})^\alpha \tag{6}$$

where the average values of $\omega'/\omega$ and $\omega/E_p$ are calculated using $F(\omega',\omega)$ and the quark momentum distribution function $q(x)$ respectively. As is immediately evident from Eq. (6) if the proton spectrum is a power law, then except near to $m_\pi c^2/2$ the gamma ray spectrum will also be a power law with the same slope. As one approaches $m_\pi c^2/2$ from above the spectrum increases, and as noted above below $m_\pi c^2/2$ there is no emission. In

Fig 2 we show our calculated $\pi^0$ decay γ ray spectrum using the spectral index α =2.75; i.e. spectral index observed for primary cosmic rays near to the earth. This is different from the spectral index (α =2.2) recommended by Liu et al. [13]; however, the shape of the γ-ray spectrum near to $m_\pi c^2/2$ is not sensitive to the exact value of spectral index. The quark momentum distribution $q(x)$ that we used is the same one we previously used in ref. 7 for positron emission, except that now the initial proton energy is interpreted as the energy of a stochastically accelerated proton. We also show as the *magenta* curve in Fig 2 our estimate for the γ ray production due to in-flight annihilation of positrons with background electrons. This in-flight positron annihilation is mainly important for nucleons with energies less than ~ 1 GeV (the cross-section for in-flight annihilation from very energetic positrons is inversely proportional to the energy of the positron). The normalization of the positron annihilation curve was chosen to be consistent with the flux of 511 KeV radiation from the galactic bulge. The important point to note is that overall there is a very sharp drop in our predicted $\pi^0$ production the spectrum below 70 MeV.

For comparison a calculation [14] for galactic $\pi^0$ decay γ-rays based on the Stecker Δ resonance model for cosmic ray $\pi^0$ production [15] is shown as the dotted curve in Fig 2. It is a matter of simple kinematics that the γ-ray spectrum due to $\pi^0$ decay is symmetric about $m_\pi c^2/2$ [16]. Of course other mechanisms for γ ray production not involving $\pi^0$ production, such as inverse Compton scattering off energetic electron, would not be expected to lead to any dramatic feature at $m_\pi c^2/2$.

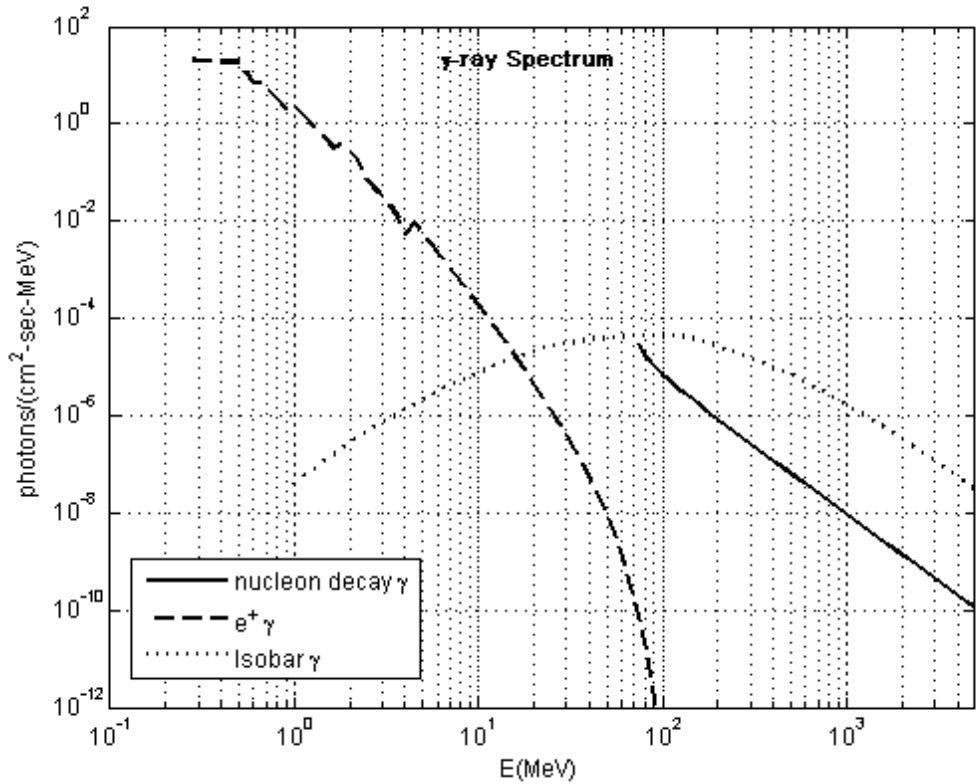

Fig. 2. The dashed curve is our calculated in-flight positron annihilation γ-ray spectrum. The solid curve is our calculated $\pi^0$ decay γ ray spectrum due to nucleon decay at the surface of a compact object. The dotted curve is the Stecker Δ isobar resonance model for cosmic ray induced galactic γ-ray production.

## 3. Conclusion

An association of 511 KeV radiation from positron annihilation with a massive compact object located near Sgr A* is circumstantial evidence for the baryon number violating processes postulated by Georgi and Glashow. These processes will produce an equally strong source flux of $\pi^0$ decay γ rays whose spectrum contains a unique feature; namely a sharp drop in the spectrum below 70 MeV. No conventional mechanism for γ ray production would contain such a feature. Observation of this spectral feature would for the first time directly confirm both baryon number violation and the failure of general relativity due to quantum effects.

**Acknowledgments**

The authors are grateful to Elliott Bloom for suggesting the idea of looking at the effects of cosmic rays hitting dark energy stars. This work performed under the auspices of the U.S. Department of Energy by Lawrence Livermore National Laboratory under Contract DE-AC52-07NA27344.